\documentclass[12pt,preprint]{aastex}
\usepackage{graphics}

\begin{document}
\title{Phase closure at 691 GHz using the Submillimeter Array}

\author{T.R. Hunter\altaffilmark{1}, A.E.T. Schinckel\altaffilmark{1}, 
A.B. Peck\altaffilmark{1},
R.D. Christensen\altaffilmark{1}, R. Blundell\altaffilmark{1},  
A. Camacho\altaffilmark{1}, F. Patt\altaffilmark{2},
K. Sakamoto\altaffilmark{1}, K.H. Young\altaffilmark{1}
}

\altaffiltext{1}{Harvard-Smithsonian Center for Astrophysics} 
\altaffiltext{2}{Academia Sinica Institute of Astronomy \& Astrophysics}

\begin{abstract}
Phase closure at 682 GHz and 691 GHz was first achieved using three
antennas of the Submillimeter Array (SMA) interferometer located on
Mauna Kea, Hawaii.  Initially, phase closure was demonstrated at
682.5~GHz on Sept. 19, 2002 using an artificial ground-based "beacon"
signal. Subsequently, astronomical detections of both Saturn and
Uranus were made at the frequency of the $^{12}$CO(6-5) transition
(691.473 GHz) on all three baselines on Sept. 22, 2002.  While the
larger planets such as Saturn are heavily resolved even on these short
baselines (25.2m, 25.2m and 16.4m), phase closure was achieved on
Uranus and Callisto.  This was the first successful experiment to
obtain phase closure in this frequency band.  The $^{12}$CO(6-5) line
was also detected towards Orion BN/KL and other Galactic sources, as
was the vibrationally-excited 658~GHz H$_2$O maser line toward evolved
stars.  We present these historic detections, as well as the first
arcsecond-scale images obtained in this frequency band.
\end{abstract}

\section{Introduction}

The Submillimeter Array (SMA) is a joint venture of the Smithsonian
Astrophysical Observatory (SAO) and the Academia Sinica Institute of
Astronomy and Astrophysics (ASIAA). The interferometer has been under
construction since the early 1990s.  First fringes at 230~GHz were
initially obtained with a single baseline at Haystack Observatory in
October 1998.  This milestone was repeated on Mauna Kea in late
September 1999.  All eight 6-meter antennas will be commissioned by
the fall of 2003.  The surface of the primary reflector is composed of
aluminum panels with a carbon fiber backup structure.  There are 24
antenna pads arranged into four tangential rings approximating
Reuleaux triangles with a maximum baseline of 508~m.  The cryostat can
accomodate up to eight SIS receiver inserts, which provides the
capability to eventually cover the entire range from 175 to 950 GHz
that is accessible from the ground.
Further description of the SMA antennas, receivers, and IF/LO 
systems can now be found in \citet{Kubo06}, \citet{Rao05}, 
\citet{Blundell04}, \citet{Hunter02}, \citet{Saito01},
\citet{Patel00}, and \citet{Moran98}.

\section{First Fringes at 650 GHz (July 2002)}

\begin{figure}[hbt]
\centering
\resizebox{16cm}{!}{\rotatebox{-90}{\includegraphics{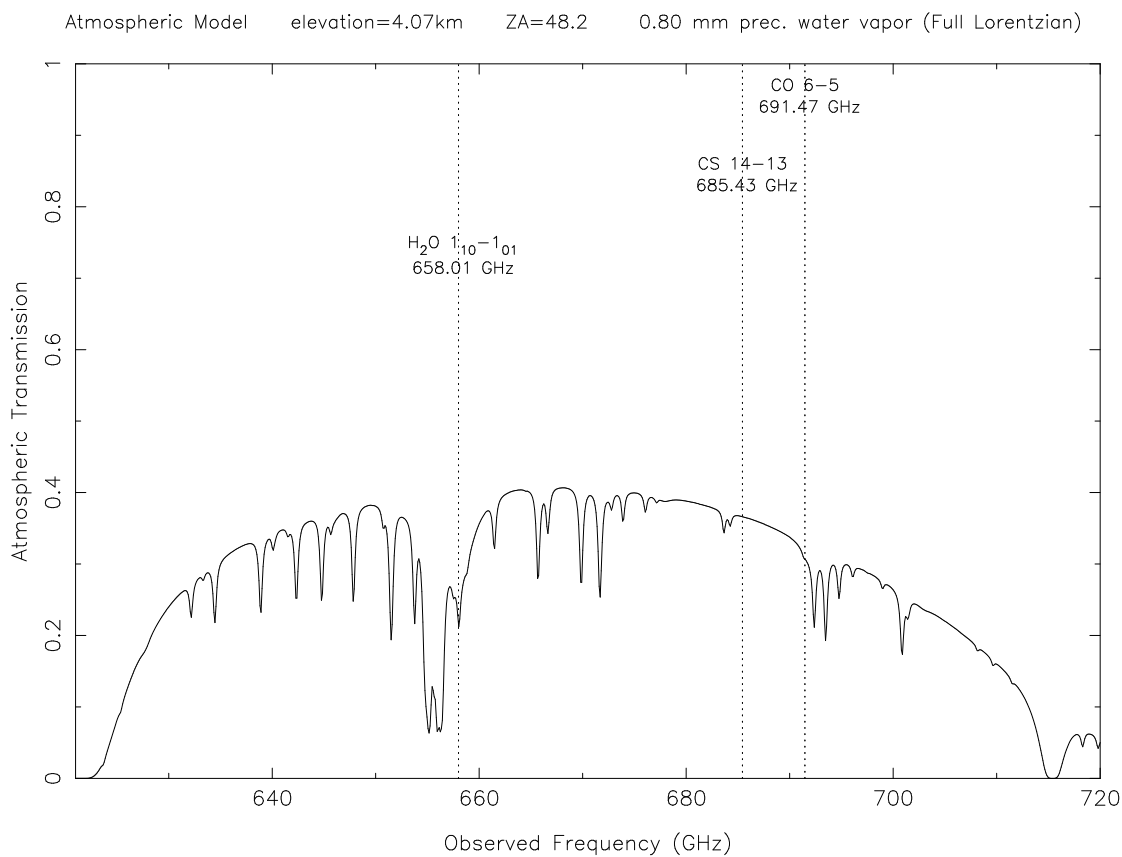}}}
\caption{The predicted opacity in the 600-700~GHz band as a function of 
frequency, at an airmass of 1.5, with a
precipitable water vapor level of 0.8mm. The line frequencies for the
spectra presented in this poster are marked. \label{fig1}}
\end{figure}

Getting the array operating successfully in the 650~GHz band was a
high initial priority for the SMA in order to enable observations at
frequencies substantially higher than any of the existing millimeter
interferometers.  A theoretical plot of the opacity in this band in
good conditions is shown in Figure~\ref{fig1}.  For some actual
measurements with a Fourier Transform Spectrometer, see
\citet{Paine00}.

\begin{figure}[htb]
\centering
\resizebox{17cm}{!}{\rotatebox{-90}{\includegraphics{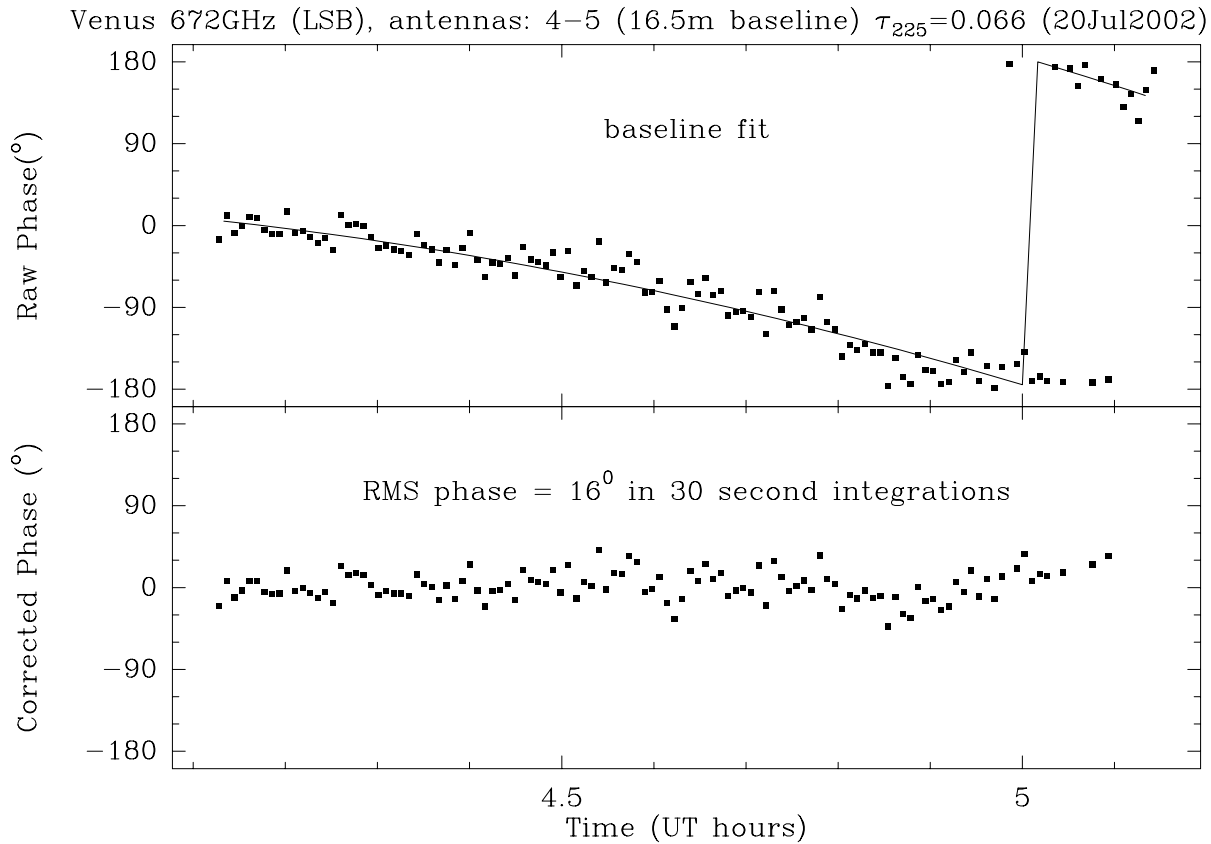}}}
\caption{These are the very first SMA fringes in the 650~GHz band.  The
baseline error has been fitted and removed, yielding a residual $16^\circ$
RMS phase variation in 30 second integrations.  This stability
bodes well for future high-frequency work on Mauna Kea. \label{fig2}}
\end{figure}

\begin{figure}[htb]
\centering
\resizebox{17cm}{!}{\rotatebox{-90}{\includegraphics{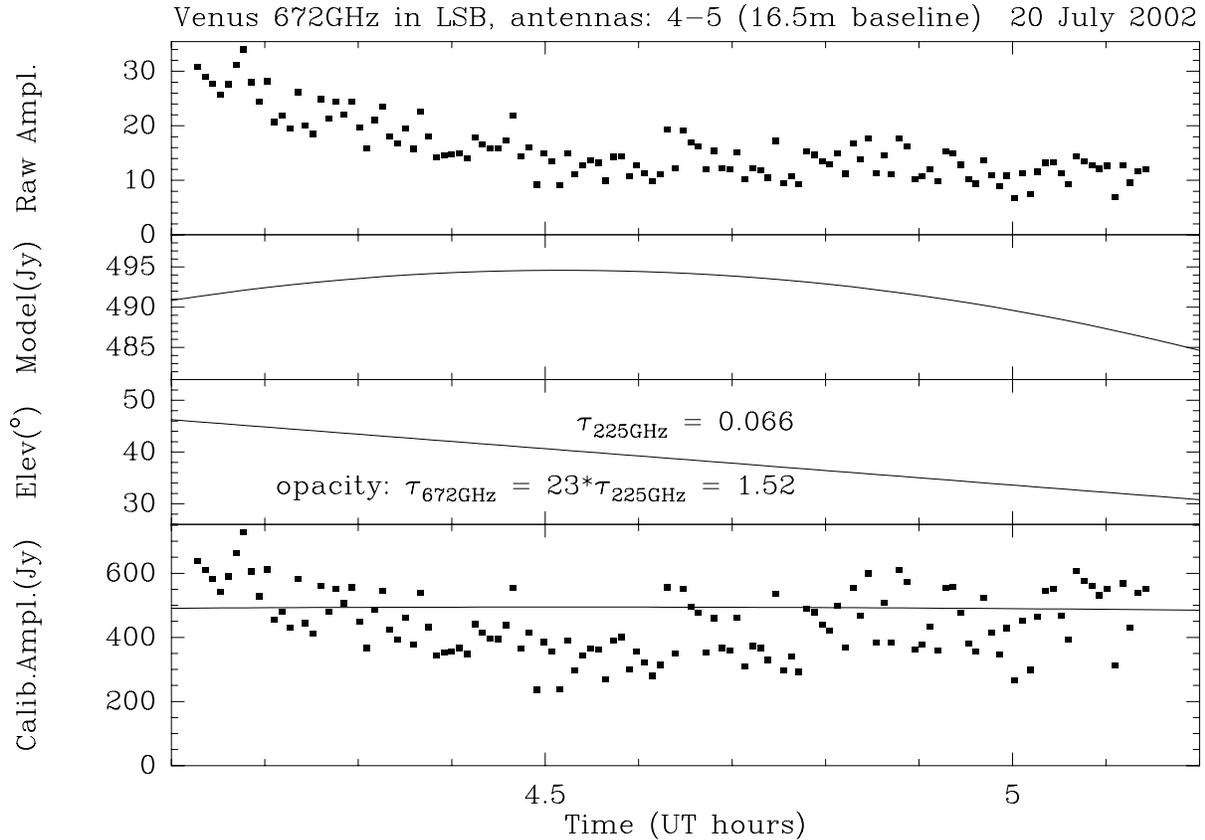}}}
\caption{In this plot, the raw correlator counts have been adjusted using  
a visibility model
for Venus and an airmass correction. This amplitude calibration
produces a fairly flat result, and has been applied to the four
star-formation regions (shown in Fig.~\ref{fig4}). \label{fig3}}
\end{figure}

On July 20, 2002, we saw first fringes on Venus with
a single 16~m NE/SW baseline at 672/682~GHz.  Although the data were
acquired during the late afternoon and early evening (usually the
worst part of the diurnal weather pattern on Mauna Kea) the RMS phase
noise was only $16^\circ$ after removing a fit to the residual error
in the baseline length (Figs.~\ref{fig2} and \ref{fig3}). 

\begin{figure}[htb]
\centering
\resizebox{17cm}{!}{\rotatebox{-90}{\includegraphics{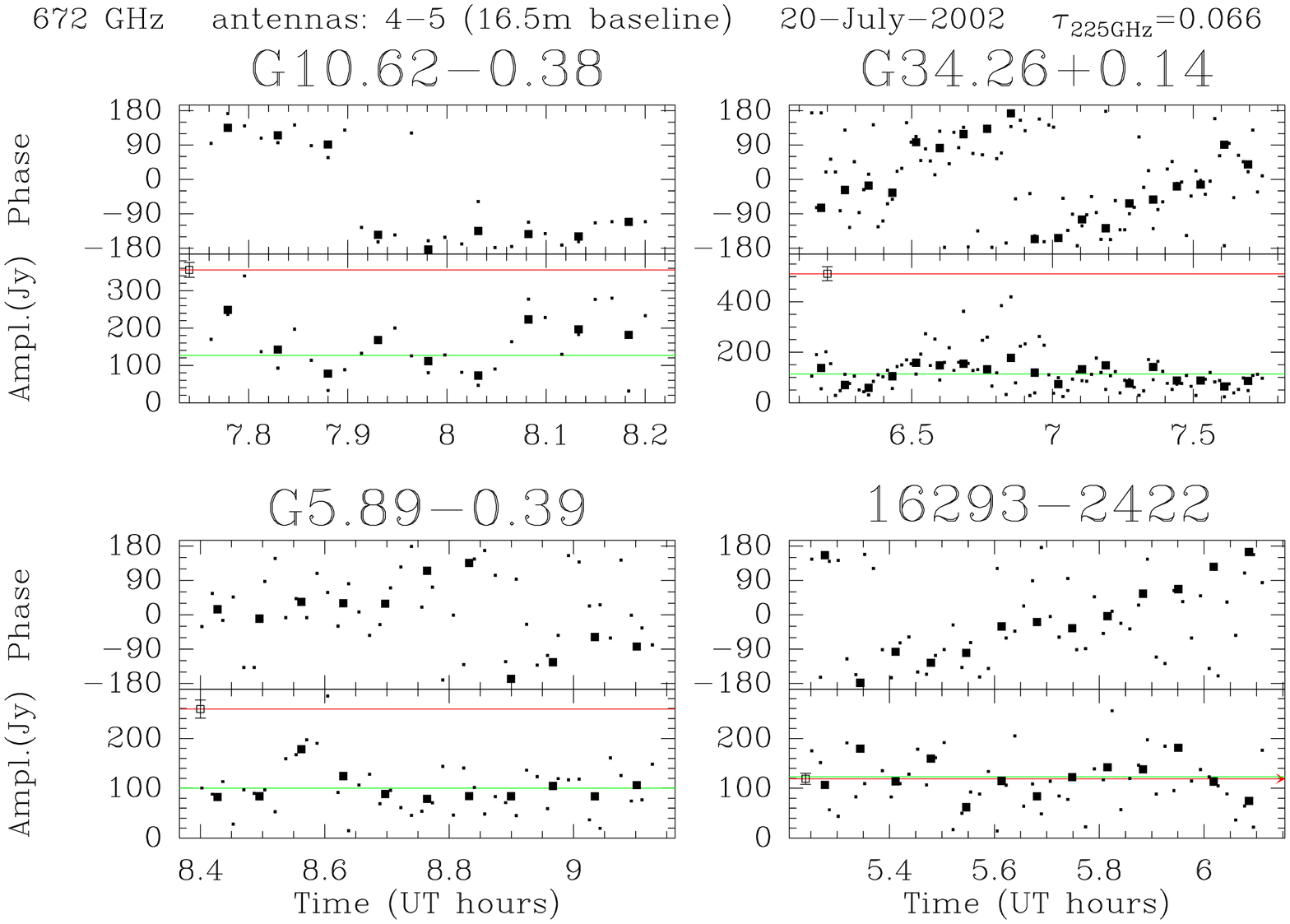}}}
\caption{
Single-baseline amplitude and phase for 4 star formation regions.
Small dots show individual integrations; large dots are 5 minute
averages.   
 The red line is the single-dish JCMT flux density \citep{Sandell94},
 and the green line is the SMA flux density.
\label{fig4}
}
\end{figure}

Several other strong dust
continuum sources were detected during the night of July 20th,
including IRAS~1629A, G10.62$-$0.38, and G34.26+0.14 (Fig~\ref{fig4}).

\section {650 GHz Phase Closure (September 2002)}

\begin{figure}[hbt]
\centering
\resizebox{16cm}{!}{\rotatebox{0}{\includegraphics{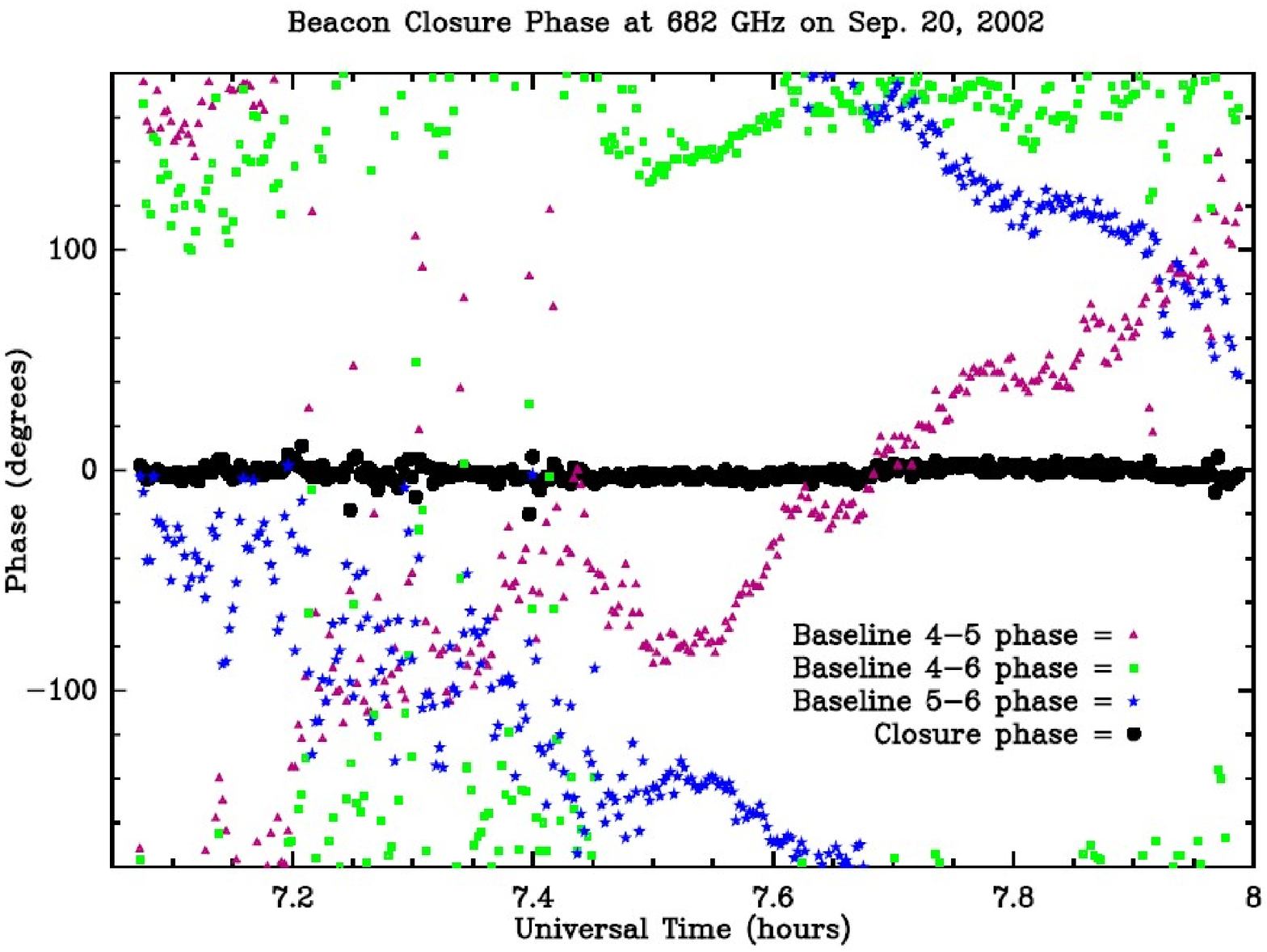}}}
\caption{ Shown are the phase of each of the three baselines, along
with the closure
phase, for the 682.5~GHz test beacon observations. These data were
taken in rather poor weather (zenith opacity $\sim 3$) shortly after
sunset.  On this night, the response of the system stabilized as 
the atmosphere cooled.
\label{fig5}
}
\end{figure}

By late summer, we had three antennas (numbers 4, 5, and 6) working in
the 650~GHz band, in a configuration with with baseline lengths up to
25.2~m. A 682.5~GHz beacon for holography had been installed on the
exterior of the nearby Subaru telescope \citep{Sridharan}, and the
first phase closure tests were performed on September 20th using this
signal (Fig~\ref{fig5}).

\begin{figure}[hbt]
\centering
\resizebox{16cm}{!}{\rotatebox{0}{\includegraphics{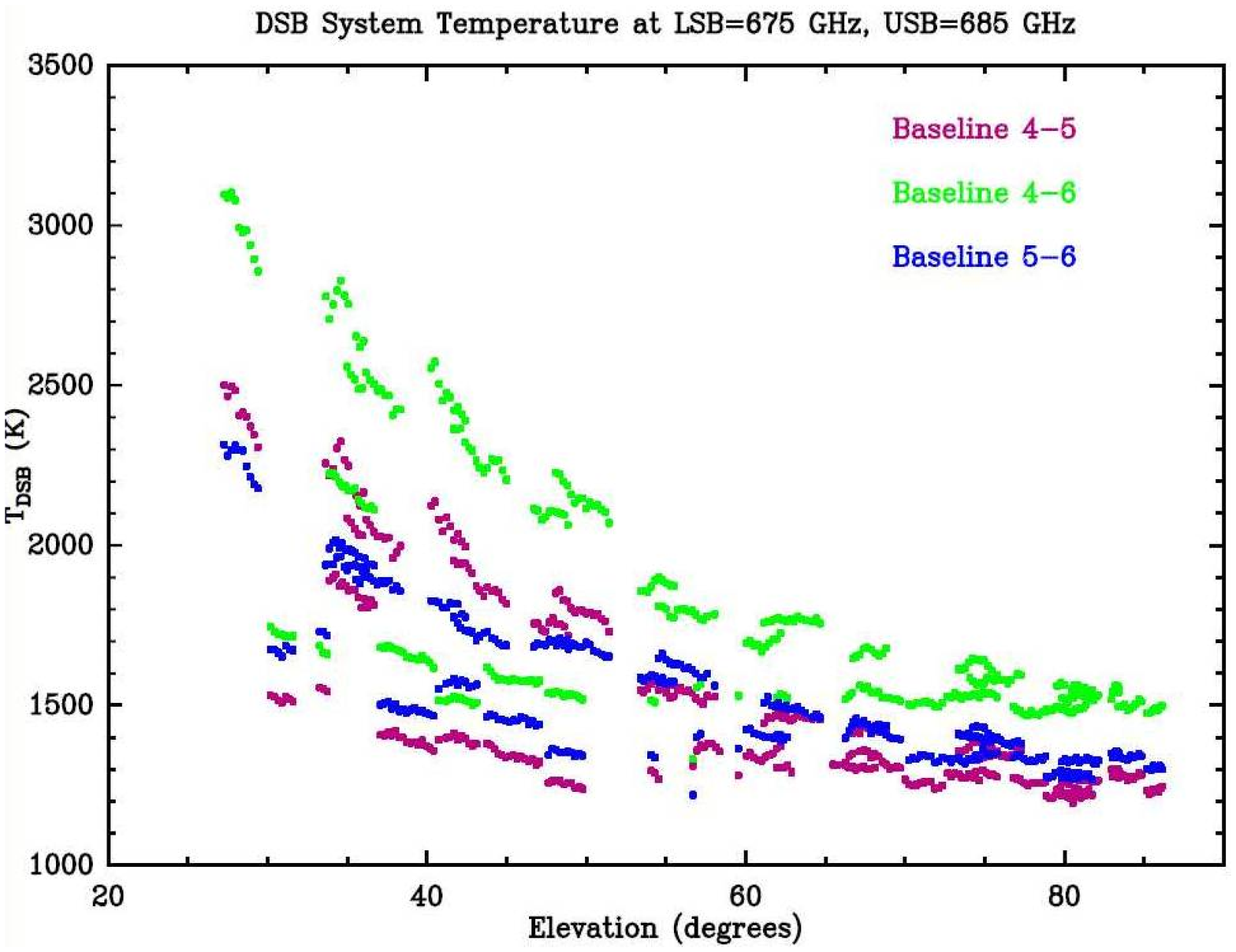}}}
\caption{Due to the high opacity, the system temperature is a strong function
of elevation and must be measured frequently, particularly for
spectral line calibration. \label{fig6}}
\end{figure}

\begin{figure}[hbt]
\centering
\resizebox{16cm}{!}{\rotatebox{0}{\includegraphics{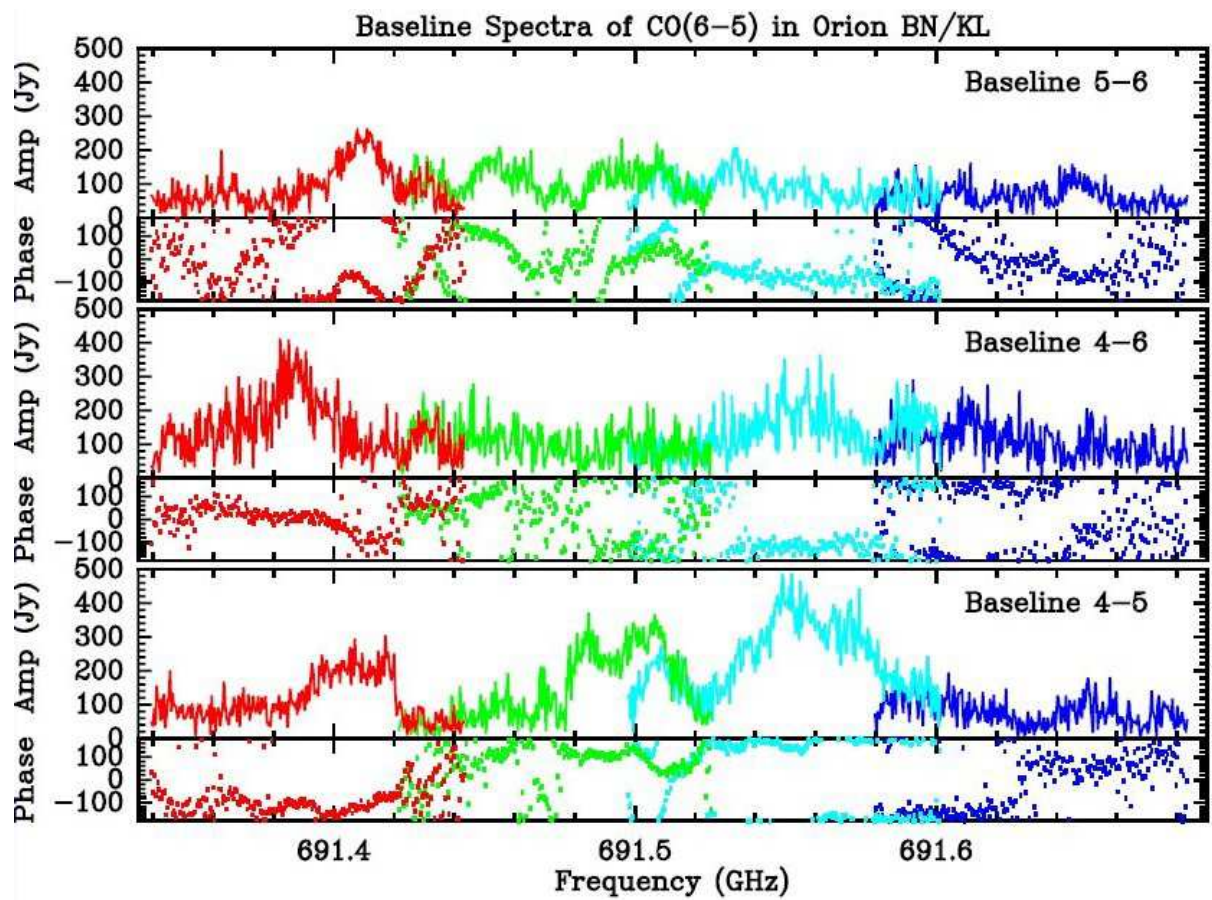}}}
\caption{
The $^{12}$CO(6-5) line in Orion spans 3/4 of the 328~MHz bandpass that was 
currently available at the time of the observation.  The bandwidth 
will be increased sixfold during 2003.
\label{fig7}
}
\end{figure}

\begin{figure}[hbt]
\centering
\resizebox{16cm}{!}{\rotatebox{-90}{\includegraphics{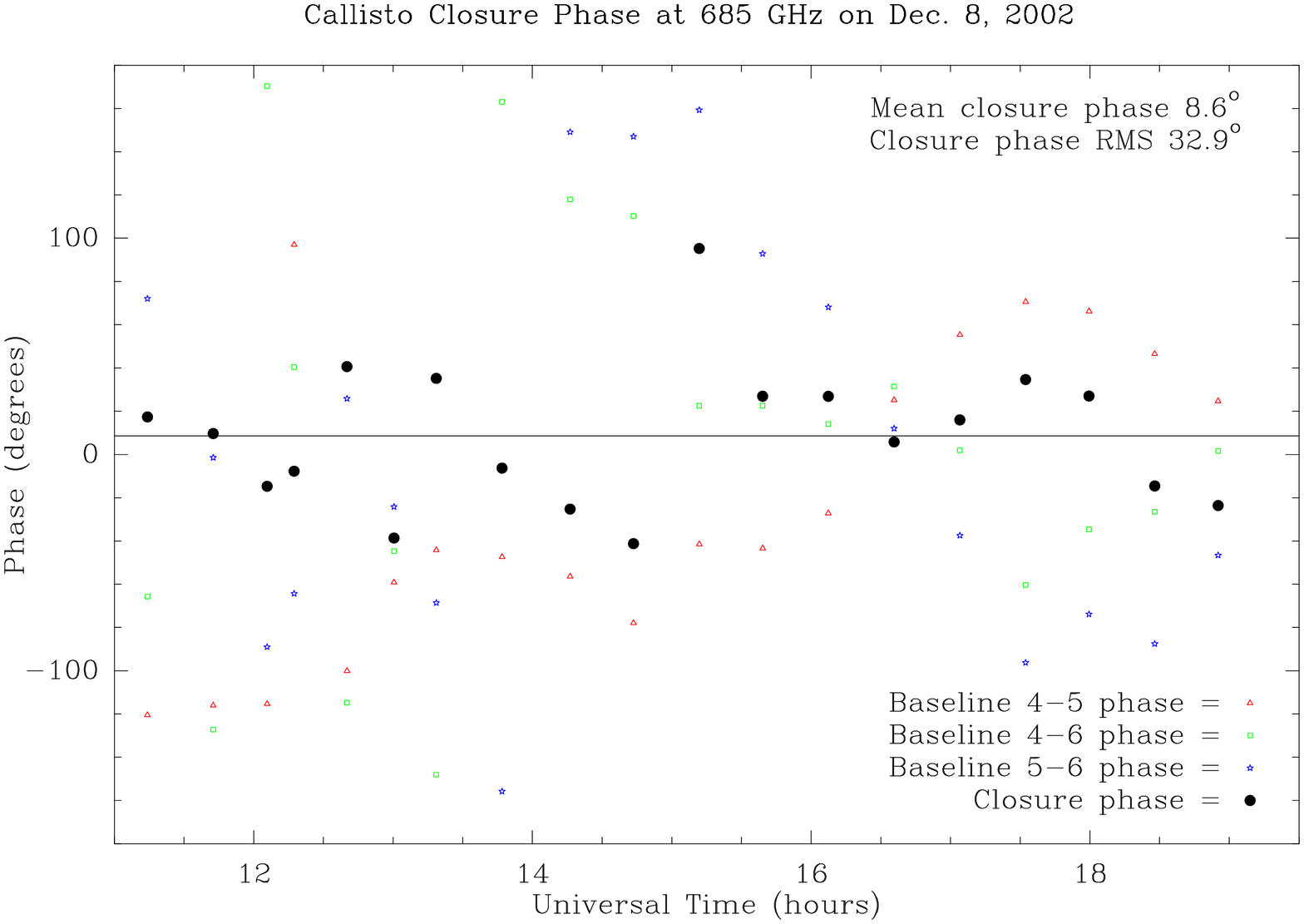}}}
\caption{
The phase closure result on Callisto is shown above.  The plotted points are 
vector averages of 10 one minute scans which were interspersed with 
observations of IRC+10$^\circ$216.
\label{fig8}
}
\end{figure}

The first convincing three baseline detections on a celestial source
were obtained two days later, with the array re-tuned to place the
$^{12}$CO(6-5) line in the USB. The zenith opacity in this frequency
band was approximately 1.2 which causes a large elevation dependence
in the system temperature (Fig~\ref{fig6}).  The phase closure on
Uranus was noisy, but the value averaged to the expected result of
zero. Saturn was easily detected on all baselines, but since it was
highly resolved, the closure phase was not constant.  Strong
$^{12}$CO(6-5) emission was detected strongly in the Orion BN/KL hot
core (Fig.~\ref{fig7}).  A convincing closure phase was also seen in
the continuum on Callisto (Fig.~\ref{fig8}).

\section {First 650 GHz Image (December 2002)}

\begin{figure}[hbt]
\centering
\resizebox{16cm}{!}{\rotatebox{0}{\includegraphics{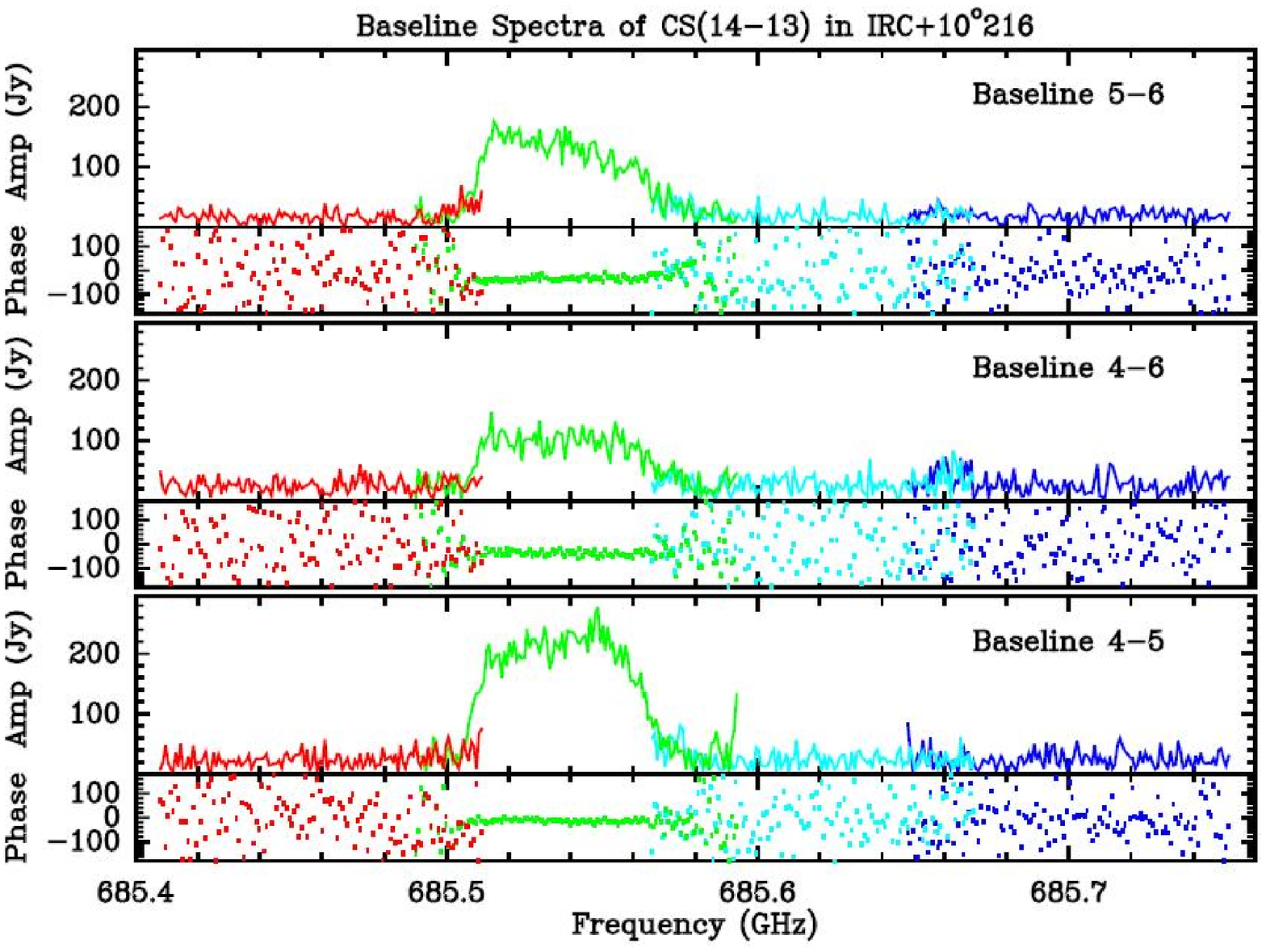}}}
\caption{
\label{fig9}
The CS(14-13) spectrum of IRC+10$^\circ$216 is shown after bandpass and
complex gain calibration.  The different colors show different, partially 
overlapping sections of the correlator.
}
\end{figure}

\begin{figure}[hbt]
\centering
\resizebox{17cm}{!}{\rotatebox{0}{\includegraphics{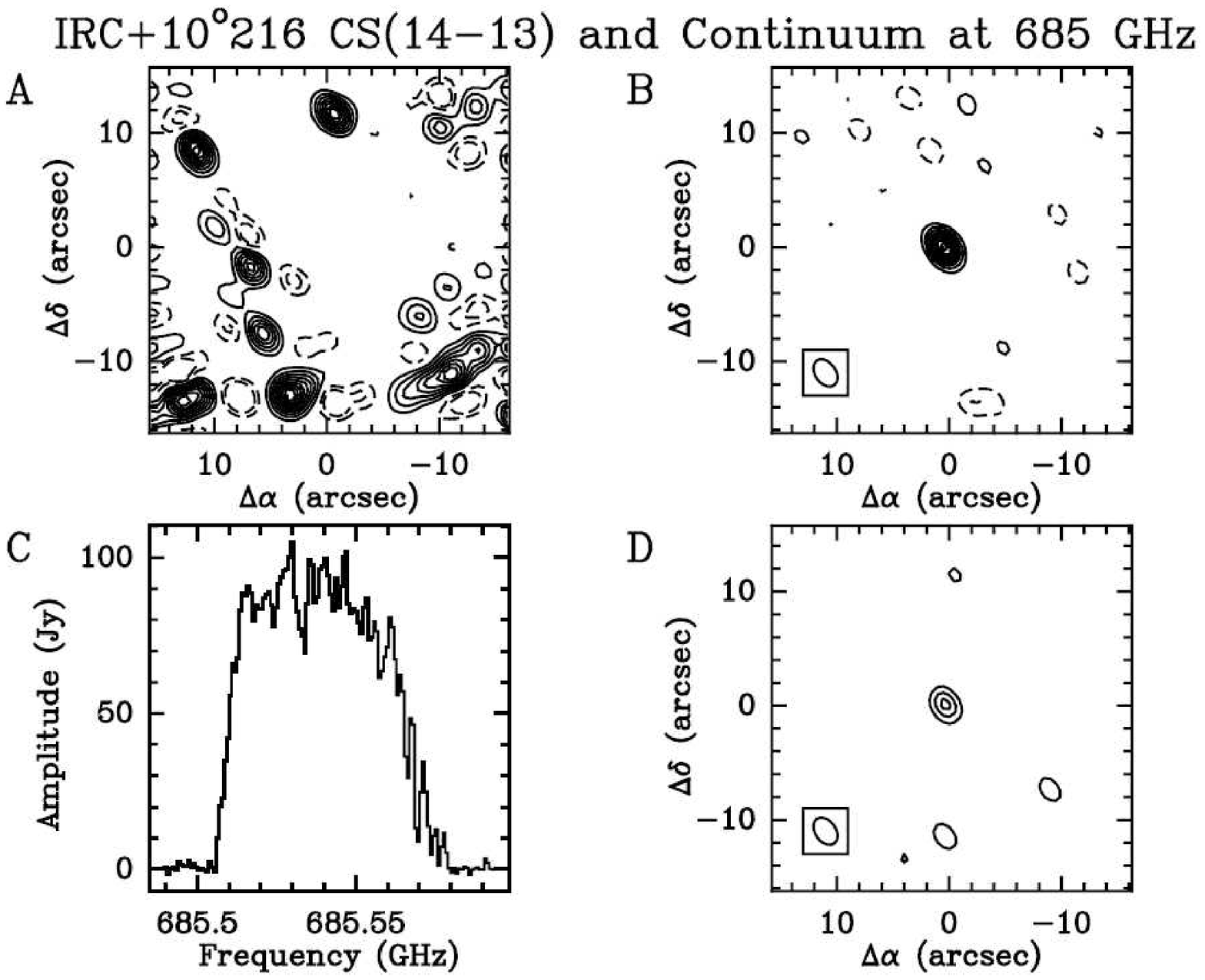}}}
\caption{
A) The CS(14-13) data without any calibration. B) Image of CS(14-13) having
used Callisto as the complex gain calibrator. C) Spectrum extracted from 
the central 
position of the calibrated image. D) Continuum image made using 
self-calibration
table derived from the line channels.  The continuum peak intensity is 
5.5~Jy/beam (see \citep{Young04}).
\label{fig10}
}
\end{figure}

One of the biggest problems for interferometric imaging in the
650~GHz band is the lack of strong point sources usable for the
derivation of the complex gain as a function of time.  Nearly all of
the quasars used at millimeter wavelengths will be too weak for use as
calibrators.  Compact thermal sources such as small planets, planetary
moons and asteroids will have to be used instead.  As a star towards
doing this, we took advantage of the close proximity of Jupiter to the
evolved star IRC+10$^\circ$216 last year.  While Jupiter itself was
far too large, the Galilean moons were only slightly resolved.  On
December 8th, we were able to detect (Fig~\ref{fig9}) make a
calibrated image of CS(14-13) at 685~GHz in IRC+10$^\circ$216 using Mars
for passband calibration and Callisto as a complex gain calibrator
(Fig.~\ref{fig10}).
CS(14-13) has a high critical density ($3.2 \times
10^8$~cm$^{-1}$) and was expected to be unresolved on our current
short baselines.\footnote{These data were later published by
\citet{Young04}.}  

\begin{figure}[hbt]
\centering
\resizebox{17cm}{!}{\rotatebox{0}{\includegraphics{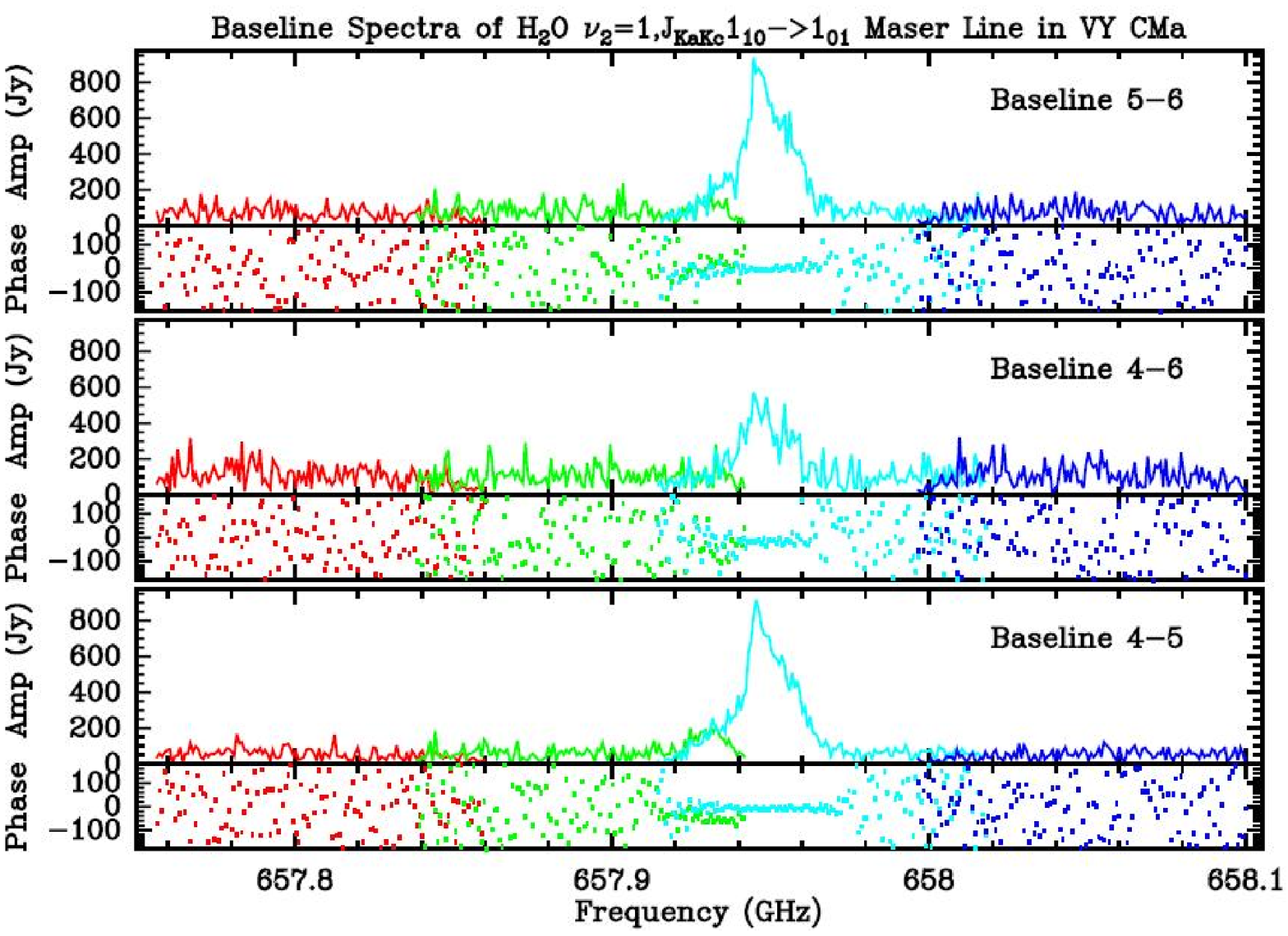}}}
\caption{ The plot above shows the very strong water maser line at
658~GHz. This line has only been detected in evolved stars
\citep{Menten95}. This line is so strong that evolved stars may serve
as gain calibrators or even pointing sources of future 
observations \citep{Hunter05}.
\label{fig11}
}
\end{figure}

On December 12th, we tuned down to 658~GHz, and
made the first interferometric observations of the strong
vibrationally-excited water line there \citep{Menten95}.  The line was
detected in VY~CMa, U~Ori and W~Hya (Fig.~\ref{fig11}).

\end{document}